# Atmospheric dispersion management in the mid-IR mode-locked oscillators

EVGENI SOROKIN,[1,*] ALEXANDER RUDENKOV,[2] VLADIMIR KALASHNIKOV,[2] AND IRINA SOROKINA[2]

[1]*Photonics Institute, Vienna University of Technology, 1040 Vienna, Austria*
[2]*Department of Physics, Norwegian University of Science and Technology, N-7491 Trondheim, Norway*

*\* Evgeni.sorokin@tuwien.ac.at*

**Abstract:** The atmospheric dispersion in the mid-infrared transparency windows presents and important albeit frequently neglected factor when developing ultra-short pulsed lasers. We show that it can amount to hundreds of fs$^2$ in 2-3 µm window with typical laser round-trip path lengths. Using the Cr:ZnS ultrashort-pulsed laser as a test-bed, we demonstrate the atmospheric dispersion influence on a femtosecond and chirped-pulse oscillator performance and show that the humidity fluctuations can be compensated by active dispersion control, greatly improving stability of mid-IR few-optical cycle laser sources. The approach can be readily extended to any ultrafast source in the mid-IR transparency windows.

## 1. Introduction

The mid-infrared (mid-IR) wavelength range, broadly defined as the range between roughly 2 and 20 µm [1] is also often called a "molecular fingerprint" region. Ultrafast sources is this wavelength region may be attractive for many applications such as sensing, bio-medicine, material processing, and science. A large progress has been made in the last decade for crystalline lasers, most notably Cr:ZnSe and Cr:ZnS [2, 3]. In particular, the range between 2 and 3 µm is characterized by the presence of the strong fundamental and overtone vibrational absorption lines of atmospheric constituent gases and vapours. Those include water vapour ($H_2O$), having the maximum around 2.6 - 2.7 µm, carbon monoxide (CO) with strong features around 2.3-2.4 µm, methane, carbon dioxide and nitrous oxide ($N_2O$) and others, having several absorption lines within this range. This spectral overlap makes the sources also prone to the adverse influence of these very molecules when they appear in the atmosphere.

For broader usage the mid-IR lasers have to be convenient in handling, not overly expensive and, most importantly, provide stable and reliable operation, be it a supercontinuum source [4, 5], frequency comb [6, 7], spectroscopic instrument [8, 9], or a seed for further amplification [10-12]. This is, however confronted by the observed dependence of device parameters on atmospheric conditions, in particular the humidity. Complete evacuation of the sources could solve the problem, at a cost of price, size, and user unfriendliness. It is important therefore to characterize and be able to counteract the influence of he naturally changing atmosphere on the operation parameters of such sources.

## 2. Air dispersion

Before proceeding we must note that the ultrafast laser is a sufficiently robust system, and it is quite stable against the losses, introduced by the narrow absorption lines of the molecular gases. The direct influence of the narrow absorption lines has been observed and well understood [13-16] to cause only additional narrowband peaks on the spectrum, which hardly affect the pulse itself. The purpose of this work is to study the influence of the dispersion, associated with the absorption lines.

There exist many experimental techniques to assess the dispersion, which include direct measurement of the group delay in the time domain and interferometric techniques in the frequency domain, and a vast literature on these methods. These techniques experience difficulties when the measured materials exhibit strong absorption, and since calculation of the second-order dispersion involves differentiation, low and hence noisy signal results in the very high uncertainty of the result.

Another way of dispersion calculation is to use the Kramers-Kronig relation from the known absorption data of the atmosphere constituents, which can be taken from e.g., HITRAN database [17-19]. The narrow rovibrational absorption lines there are represented by the Voigt contour, which is a convolution of the pressure broadening (Lorentzian) and Doppler broadening (Gaussian). Summing up all lines provides the absorption spectrum in a very broad range from the UV to far infrared for any given atmosphere composition, temperature and pressure. Taking this spectrum in the frequency domain as the imaginary part of the complex index of refraction allows calculating its real part

$$n(\omega) = 1 + \frac{c}{\pi} v.p. \int \frac{\alpha(\omega')}{\omega'^2 - \omega^2} d\omega'. \tag{1}$$

The propagation constant $\beta(\omega) = \omega n(\omega)/c$ can then be numerically differentiated to obtain the the group velocity dispersion (GVD) $\beta_2 = d^2\beta(\omega)/d\omega^2$. At normal atmospheric conditions, where the pressure broadening dominates, a standard procedure is to use Lorentzian shapes, where analytical expressions can be used to simplify calculations [20].

Direct calculation according to expression (1) for gases results in a wildly oscillating GVD curve (Fig. 1). This is because the second derivative of the narrow line with characteristic width $\Delta\omega$ scales as $\Delta\omega^{-2}$ and for the typical width of the atmospheric lines at normal condition of ~ 0.1 cm$^{-1}$ even a very weak absorption line produces oscillations with ps$^2$/m amplitude and more. For the ultrashort pulse propagation this kind of GVD is not relevant, because physically the narrow absorption feature only results in a long tail (free induction decay) behind the pulse with duration ~ $\Delta\omega^{-1}$ being many tens of picoseconds and its overlap with the pulse is negligible. However, each absorption line does contribute to a slowly varying refractive index curve far beyond the line width, which will result in a GVD. It becomes necessary to correctly remove the narrowband features from the calculated curve in Fig. 1.

We have tested and compared few approaches to retrieve the slowly varying part of the GVD: 1) brute-force smoothing of the GDD curve, 2) simulation of a time-domain white-light interferometer, 3) simulation of an autocorrelator, 4) local line exclusion, 5) exclusion of the line center vicinity, and 6) computing of the group delay of the pulse, defined either as a delay of its intensity maximum or the first moment. The details of these procedures are described in the Supplement, here we will only summarize the common features. First, all methods provide compatible results in the transparency windows, where only small absorption is present, while strongly differing in the areas with high absorption, such as e.g. water vapour band around 2.7 μm. Second, the results critically depend on the spectral width parameter: window width for the for the smoothing, pulse spectral width for the white-light interferometer, autocorrelator and group delay calculation, and the width of the excluded area for the line exclusion or line center exclusion techniques.

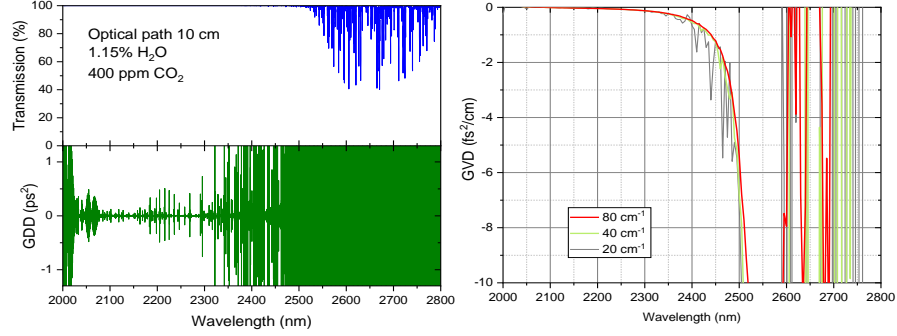

Fig. 1. (a) Transmission of the atmosphere and corresponding GDD for the 10 cm optical path at 40% relative humidity (r.h.) at 296 K. Note the scale of the GDD, which is in *picoseconds* squared! (b) The GVD data processed by the exclusion of the local lines method with different width of the excluding window.

For practical use, one should also establish the dependence of the dispersion on the atmosphere parameters. It follows from the basic principles, that the real part of the refractive index, associated with a narrow absorption line at $\omega_0$, has broad wings which decay asymptotically as $(\omega_0 - \omega)^{-1}$ [21]:

$$n(\omega) \approx 1 + \frac{c}{2\pi\omega_0(\omega_0 - \omega)} \int \alpha(\omega')\, d\omega'. \qquad (2)$$

Validity of this approximation can be easily seen analytically for the Lorentzian lineshape, and the integral in this expression is independent of the exact shape and broadening factor of the absorption line. Thus, the contribution of a line to the slowly varying dispersion does not depend directly on pressure or temperature, but only on its integral cross-section and the volume concentration of absorbing molecules in the atmosphere. Also, if the temperature is within the normal laboratory ranges, the cross-section distribution of the rotational components does not vary much and can be taken at HITRAN reference temperature of 296 K. This allows using the tabulated GVD results as is scaled by the molecule concentration.

## 3. Experimental verification

To verify our model calculation we have constructed a Kerr-lens modelocked (KLM) ultrashort-pulse oscillator based on a Cr:ZnS laser (Fig. 2), with pulse repetition rate 69 MHz and round-trip air path $L = 433$ cm. We have detailed dispersion data of all materials and of all mirrors. The mirror group delay dispersion (GDD) can be calculated from the coating design, but to allow for the layer thickness error during manufacturing, the layer thicknesses have been corrected by refining the design using the measured transmission curve. The YAG wedges allowed fine and controllable adjustment of the GDD.

The oscillator has been adjusted to operate in the vicinity of the zero GDD so that it was possible to switch the round-trip GDD from normal to anomalous by changing the wedge insertion or adding the calcium fluoride Brewster plate. With anomalous round-trip GDD the laser operated in the femtosecond regime with $\tau = 30\text{-}60$ fs and 150-300 nm spectral width at half-maximum. Undoer normal round-trip GDD the laser operated as a chirped-pulse oscillator (CPO [22], also called dissipative soliton in the literature [23]) with $\tau = 2\text{-}3$ ps and 120-180 nm spectral width. To exclude thermal effects, we operated the oscillator at relatively low output power of 100 mW per OC bounce, corresponding to 14.4 nJ intracavity energy at 69 MHz repetition rate. Since both the soliton and CPO regimes are subject to respective area theorems [24, 25] connecting their duration and energy, we made sure to maintain exactly the same intracavity energy in every data series.

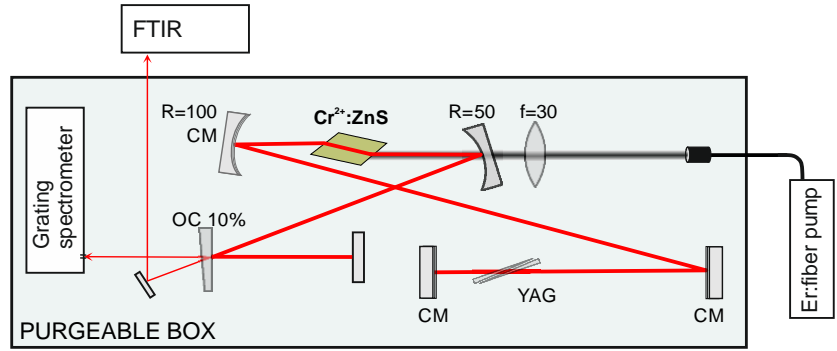

Fig. 2. Experimental setup. CM: the high reflecting mirrors which could be substituted for chirped mirrors. OC: output coupler. YAG: a Brewster pair of the 3° wedges made of YAG.

To control the atmosphere, the laser has been put in a box, which was purged with the dry nitrogen, with the spectrometer inside the box. We could continuously measure the humidity inside the box, however, we found that this was a very inaccurate measurement. The humidity sensor was quite slow and measured the humidity only at its own position in the box. We therefore assessed the water vapour concentration by measuring the relative amplitude of the feature at 2434 nm on a soliton spectrum, which is proportional to the total absorption of the corresponding water line [13]. This is a much more relevant data, as it is immediate and shows the average water molecule concentration along the actual beam path.

*Soliton regime*

To obtain the possibly short pulses one would strive to provide small and possibly flat negative GDD over the whole expected range. Consequently, the oscillator becomes quite sensitive to the dispersion changes due to the humidity fluctuations, as illustrated by Fig. 3. The effect of the humidity change can be quite strong but very different. It is nearly impossible to characterize it because the exact round-trip GDD curve has inevitable modulation from the chirped mirrors.

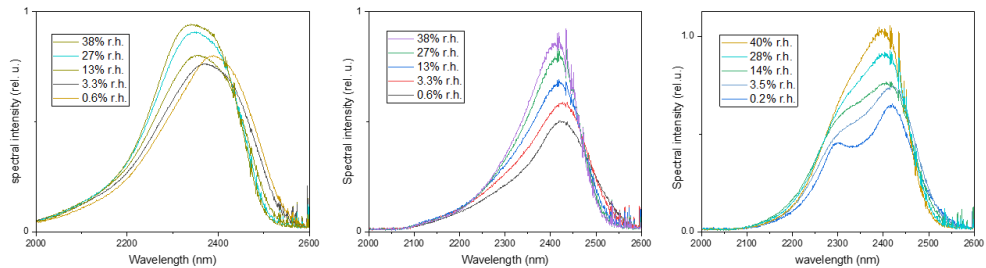

Fig. 3. Effect of humidity change on the oscillator spectra. The oscillators had different combinations of mirrors: scheme as in Fig. Yy with two chirped mirrors (left graph), same scheme but with a single bounce from a 20% output coupler and different YAG wedge insertions (middle and right graphs).

Adding a 3 mm CaF$_2$ Brewster plate turns the oscillator to a longer-pulse regime with $\tau = 50 - 55$ fs and strong dispersive-wave feature at 2060 nm due to the uncompensated third-order dispersion. In this regime, purging the box produces results only in a slight shift of the spectrum towards the longer wavelengths (Fig 4), following a certain spectral position with $\beta_2 \cdot L \approx -400$ fs$^2$, to fulfill the area theorem condition. In this setup with dominating third-order dispersion and spectrum lying mostly within the small-dispersion range, humidity change only modifies the GDD value at the pulse spectral location.

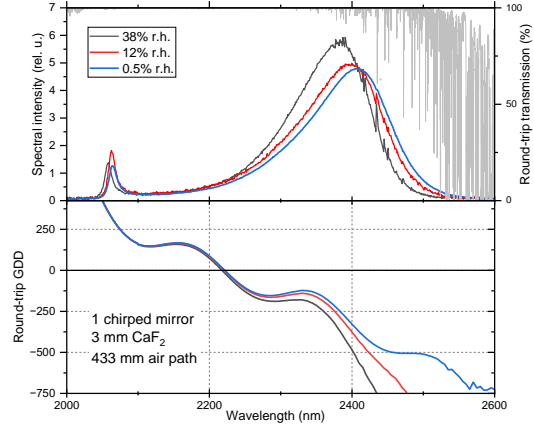

Fig 4: Spectra and GDD of the long-pulse oscillator setup. The gray star marks the water absorption line, which was used as a humidity sensor.

*CPO regime*

The chirped-pulse in solid-state lasers is known to be very sensitive to the dispersion[26] and even very small changes result in the visible deformation of the spectrum. In addition to that, the CPO spectrum is cut at the edges, so that it interacts only with a limited and clearly identifiable range of the GDD curve. In our setup, this range was well in the water-free window so that there were no additional sources of losses at the wings. A typical behaviour of a CPO oscillator under purging is shown in Fig. 5 a).

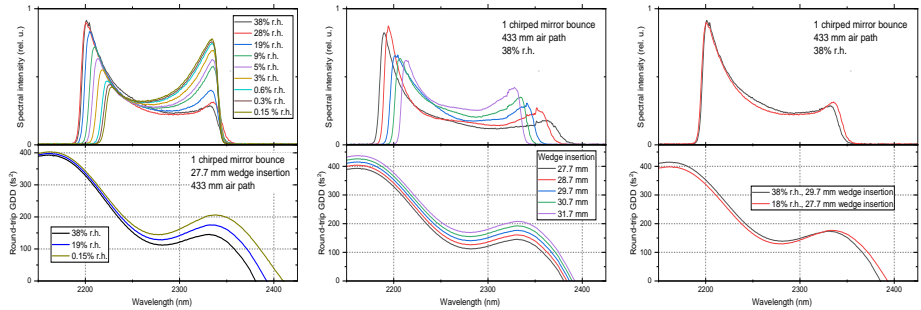

Fig. 5. (a) CPO oscillator spectra under nitrogen purging. (b) CPO oscillator spectra at open air with different YAG wedge insertions. (c) Approximate compensation of higher humidity by increased YAG Wedge insertion. For all graphs, the lower plot shows the calculated round-trip GDD.

At the same time, changing the insertion of the YAG wedges also changes the GDD, as shown in Fig. 5 b). By manipulating the wedge insertion and purging it is possible to approximately compensate both effects, so that the spectra become nearly identical (Fig. 5 c).

As one can see from the GDD curves, the dispersion increase due to the wedges insertion nearly compensates the atmospheric GDD decrease due to the purging. This as a direct experimental confirmation for dispersion calculation procedure, allowing one to design a humidity compensator, that would use the independent humidity, temperature, and pressure gauges to control the dispersion in real time, thus greatly improving stability of the oscillators.

## 4. Conclusion

The atmospheric influence in the MID-IR is significant even in the nominally transparent wavelength regions. The reason is less the absorption, but rather the dispersion influence, which is comparable to that of the net GDD of a typical oscillator, operating around 100 MHz rep. rate. From the atmospheric constituents, the air humidity makes the largest contribution in the wavelength range below 3.5 µm, while CO2 introduces much stronger effect around its 4.2 µm absorption band. If complete evacuation is not desirable, it is possible to use the finely adjustable dispersive elements, such as wedge pairs, to compensate for the natural humidity and pressure fluctuations. The exact amount of required compensation can be calculated with the HITRAN data and used e.g. in an automated control scheme with humidity, temperature, and pressure gauges.

**Funding.** The work is supported by the Norwegian Research Council projects #: 326503 (MIR), 303347 (UNLOCK).

**Disclosures.** The authors declare no conflicts of interest.